# Reliability of Critical Infrastructure Networks: Challenges

**Konstantin M. Zuev, Ph.D.[1] and Michael Beer, Prof. Dr.-Ing.[2,3,4]**

[1]Department of Computing and Mathematical Sciences, California Institute of Technology, 1200 E. California Blvd., Pasadena, CA, USA; e-mail: kostia@caltech.edu
[2]Institute for Risk and Reliability, Leibniz Universität Hannover, Callinstraße 34, 30167 Hannover, Germany; e-mail: beer@irz.uni-hannover.de
[3]Institute for Risk and Uncertainty, University of Liverpool, UK
[4]International Joint Research Center for Engineering Reliability and Stochastic Mechanics, Tongji University, China

## ABSTRACT

Critical infrastructures form a technological skeleton of our world by providing us with water, food, electricity, gas, transportation, communication, banking, and finance. Moreover, as urban population increases, the role of infrastructures become more vital. In this paper, we adopt a network perspective and discuss the ever growing need for fundamental interdisciplinary study of critical infrastructure networks, efficient methods for estimating their reliability, and cost-effective strategies for enhancing their resiliency. We also highlight some of the main challenges arising on this way, including cascading failures, feedback loops, and cross-sector interdependencies.

## 1. INTRODUCTION

Critical infrastructure networks, such as transportation systems (roads, rails, and airlines), electric power grids, natural gas and petroleum networks, water distribution networks, cellular grids, and the internet, are tightly interwoven into the fabric of the modern world. These complex distributed systems ensure the functioning of our society by providing us with services critical for everyday life, such as water, food, energy, banking, and finance. Moreover, they facilitate transport-dependent economic activities, and make communication and access to information fast and efficient. In a sense, critical infrastructures form a technological skeleton of our civilization.

About ten years ago, we have reached a "tipping" point when, for the first time in human history, more than half of the world's population was living in cities. For more developed regions, such as North America and Europe, this happened even earlier. Figure 1 shows the dynamics of the percentage of total population living in urban areas for different regions between 1967 and 2015. Furthermore, it is projected that by 2050, 66% of the world's population will be urban (United Nations 2014). As urban conglomerations are growing, the dependence of our society on critical infrastructure networks, spanning cities, countries, and even continents, is constantly increasing. This makes the study of critical infrastructure networks, the enhancement of their resiliency, and the quantification of the associated uncertainties one of the most challenging and important problems of modern engineering.

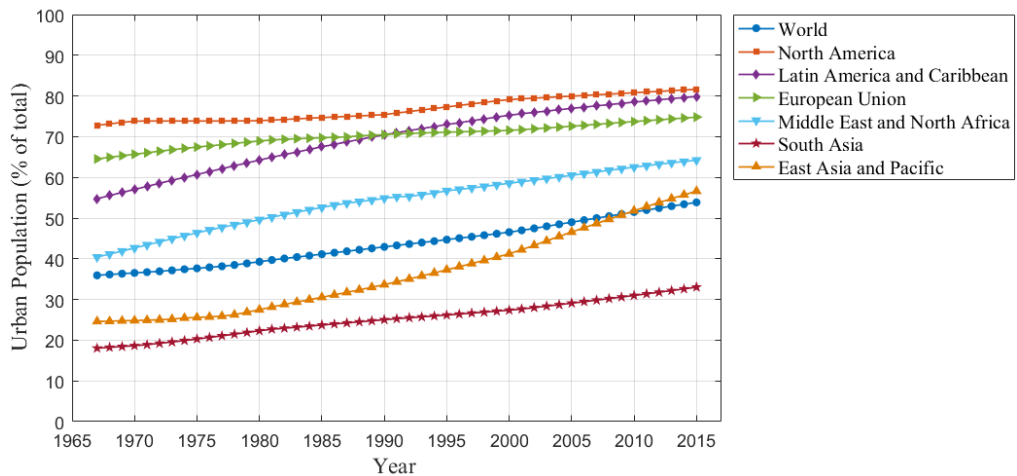

**Figure 1. Percentage of population living in urban areas by region, 1967-2015. Data from database: World Development Indicators,** <http://databank.worldbank.org/>.

Improving the resilience of (especially aging) infrastructures is absolutely necessary for avoiding failures such as the one happened in 2010 in San Bruno (a suburb of San Francisco, CA), when a 30-inch diameter natural gas pipeline (installed in 1956) exploded, killing 8 people, injuring 58, and destroying 38 homes (Lagos et al. 2010; NTSB 2011). Because of the strong shaking caused by the explosion, some of the local residents, first responders, and news media thought initially that it was an earthquake. It took nearly an hour to find out that it was actually a natural gas pipeline explosion. Although this failure was truly devastating, it was still a "local" failure in the sense that it did not propagate and did not cause any other failures. But in principle the consequences of this local failure could have been even worse.

It is well known that the network of natural gas pipelines and electric power grid are strongly interconnected, since pipelines provide fuel for electrical generators and the grid provides power for compressors, storage, and control systems in the gas network. This coupling creates a wealth of potential scenarios for failure propagation, one of which is schematically illustrated in Figure 2: the pipeline explosion (denoted by the fire sign) causes an electrical generator failure in the grid, which triggers a sequences of cascading power outages (power grids are prone to cascading failures), which, in turn, leads to substantial losses of natural gas production. Electric power and other infrastructure problems in California provide many examples of this kind of interdependencies and feedback loops (Rinaldi et al. 2001).

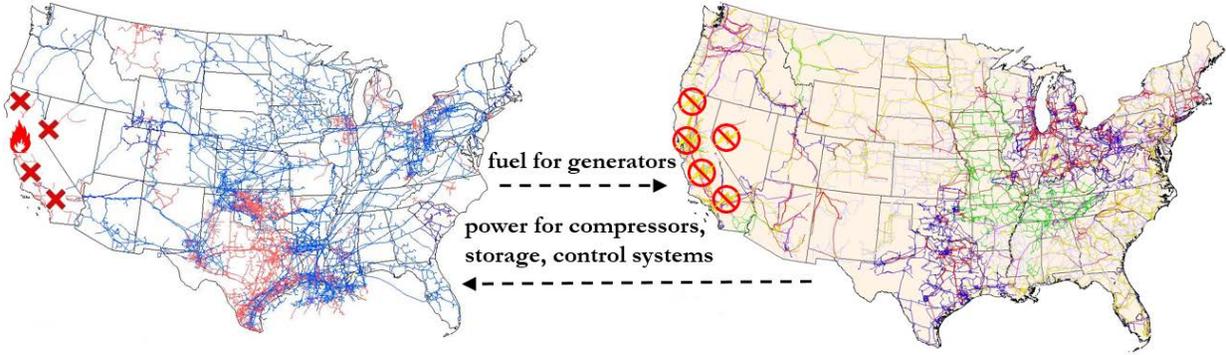

**Figure 2. Failure propagation in coupled infrastructure networks: U.S. natural gas pipeline network (on the left, source: Energy Information Administration) and U.S. power grid (on the right, source: <https://commons.wikimedia.org/wiki/File:UnitedStatesPowerGrid.jpg>).**

In this paper, we adopt a network view on critical infrastructures and discuss how to estimate their reliability and what challenges need to be overcome to make this accurately and efficiently. The rest of the paper is organized as follows. In Section 2, we formulate a general network reliability problem and explain why it is computationally very challenging. In Section 3, we briefly describe a recently introduced network reliability method, and, taking this method as an example, we focus on the challenges in Section 4. Section 5 concludes with a brief summary.

## 2. NETWORK RELIABILITY PROBLEM

Suppose that an infrastructure network is modeled as a graph with $n$ nodes and $m$ links. Assume that all nodes are absolutely reliable and don't fail, but links may fail. The network state is then represented by a vector $s = (s_1, \ldots, s_m)$, where $s_i = 1$ if the $i^{th}$ link is fully operational, $s_i = 0$ if it is fully failed, and $0 < s_i < 1$ if it is partially operational. Geometrically, the network state space, $S = \{s = (s_1, \ldots, s_m), 0 \leq s_i \leq 1\}$ is thus an $m$-dimensional hypercube, where vertices $(1, \ldots, 1)$ and $(0, \ldots, 0)$ correspond to an ideal state and complete failure state, respectively.

Next, let $f(s)$ be a probability distribution function (PDF) on the network state space $S$ which provides a probability model for occurrence of different states. In applications, the PDF $f(s)$ is constructed based on expert knowledge and available data. Also, let $u(s)$ be a performance function that quantifies the degree to which the network provides the required service (Ghosn 2016). Here, $u(s)$ is interpreted as a utility function, meaning that higher values of $u(s)$ correspond to better performance. Finally, let's define a failure domain $F \subset S$ as a collection of network states with performance below a certain critical threshold $u^*$,

$$F = \{s : u(s) < u^*\}. \qquad (1)$$

The network reliability problem is then to estimate the probability of failure $p_F$ which is given by the following integral:

$$p_F = \int_F f(s)ds = \int_{R^m} f(s)I_F(s)ds, \tag{2}$$

where $I_F(s)$ is the indicator function of the failure domain $F$, i.e. $I_F(s) = 1$ if $s \in F$ and $I_F(s) = 0$ if $s \notin F$. This problem is of fundamental importance for both risk assessment of existing technological networks and design of future infrastructures.

There exist a few factors which make estimating the failure probability (2) challenging. First, the number of links $m$ is very large. For example, the U.S. Western States power grid has more than 6,000 links (Watts and Strogatz 1998) and the California road network has more than 2.5 million links, i.e. roads connecting intersections and endpoints (Leskovec et al. 2009). Therefore, the failure domain (1) is high-dimensional and numerical integration methods are simply computationally infeasible. Next, the probability of failure $p_F$ is very small, since the network failure is a relatively rare event. This means that if we want to employ the Monte Carlo method (Robert and Casella 2004; Zio 2013), which is a standard general method for high-dimensional integration, then, to get an accurate estimate of $p_F$, we need to generate a large number of samples. This stems from the fact that the coefficient of variation (COV) of the Monte Carlo estimate is approximately $\delta \approx 1/\sqrt{Np_F}$, where $N$ the total number of samples (e.g. Beck and Zuev 2017). Finally, computing the performance function $u(s)$ is typically time-consuming due to the complexity and size of the network. As a result, the computational effort for evaluating the integrand $f(s)I_F(s)$ in (2), involving checking whether $s$ is a failure state, is significant, which makes the use of a large number of Monte Carlo samples computationally prohibitive.

Estimating network reliability is thus indeed a very challenging problem, and we are still at the very beginning of developing efficient practical methods for solving this problem. In recent years, several promising methods have been developed for network reliability, resilience, and risk analysis. Liu and Li (2012) introduced an analytical method, based on the minimal cut recursive decomposition algorithm, for evaluating the network connectivity reliability. Fang and Zio (2013) proposed a general framework for multi-scaled representation of critical infrastructure networks, which is based on a recursive unsupervised spectral clustering method, and showed how this representation can be used for network reliability analysis at different scales. Gómez et al. (2014) used agent-based modeling for developing an efficient integrative methodology for risk assessment and management of infrastructure networks.

In what follows, we focus our attention on a recently proposed stochastic simulation method for network reliability estimation (Zuev et al. 2015), which is based on the Subset Simulation method, originally introduced by Au and Beck (2001) for estimating small failure probabilities of complex structural systems, such as buildings and bridges, subject to earthquake excitations. This focus on simulation is driven the desire for a high general applicability of the approaches for estimating network reliability. In the next two sections, we briefly describe this method and highlight some of the main challenges that need to be overcome to make it and other proposed network reliability methods accurate and efficient in real-world applications.

## 3. SUBSET SIMULATION FOR NETWORK RELIABILITY ESTIMATION

The key idea behind Subset Simulation is to decompose a very small failure probability into a product of larger conditional probabilities, each of which can be efficiently estimated by a Monte-Carlo-like method. This is achieved by considering a sequence of nested subsets, called intermediate failure domains, that starts from the entire network state space $S$ and finishes at the target failure domain $F$,

$$S = F_0 \supset F_1 \supset \cdots \supset F_{L-1} \supset F_L = F. \tag{3}$$

Each intermediate failure domain $F_i$ in (3) is defined similarly to the target failure domain in (1) by relaxing the value of the critical threshold $u^*$:

$$F_i = \{s : u(s) < u_i^*\}, \quad \text{where} \quad u^* = u_L^* < u_{L-1}^* < \cdots < u_1^*. \tag{4}$$

Given subsets (3), the failure probability can be written as a product of conditional probabilities:

$$p_F = P(F) = P(F_1)P(F_2|F_1) \ldots P(F_L|F_{L-1}). \tag{5}$$

By choosing the intermediate thresholds $u_i$ appropriately, one can make all conditional probabilities $P(F_i|F_{i-1})$ in (5) large enough, so that the corresponding conditional events $F_i|F_{i-1}$ are not rare and can be estimated efficiently by sampling.

The first factor, $P(F_1)$, can be in fact efficiently estimated by direct Monte Carlo, provided that $u_1^*$ is sufficiently large. To estimate $P(F_i|F_{i-1})$ for $i \geq 2$, one should be able to sample from the conditional distribution $f(s|F_{i-1})$, which is, in general, a nontrivial task. In Subset Simulation, this is achieved by using the Modified Metropolis algorithm (MMA) (Au and Beck 2001; Zuev and Katafygiotis 2011), which is a Markov chain Monte Carlo (MCMC) algorithm (Liu 2001; Robert and Casella 2004) specifically tailored for sampling from conditional distributions in high dimensions. The main difference between MMA and the original Metropolis algorithm (MA) (Metropolis et al. 1953) is in the way the candidate states of Markov chains are generated. Unlike MA, where a high-dimensional proposal distribution is used to generate a candidate state, MMA generates a candidate state component-wise using a sequence of univariate proposal distributions. It was shown in (Au and Beck 2001) that MMA outperforms MA in high dimensions, provided that random variables $s_1, \ldots, s_m$ are independent. A geometric explanation as to why this happens, based on the properties of multivariate normal distribution in high-dimensions, is provided in (Katafygiotis and Zuev 2008). For more details on MMA, the reader is referred to (Zuev et al. 2012), where both the algorithm and its practical implementation are discussed in detail.

The efficiency of Subset Simulation strongly depends on the choice of intermediate critical thresholds $u_1^*, \ldots, u_{L-1}^*$, since they control the values of conditional probabilities $P(F_i|F_{i-1})$ in (5). It is difficult to find the optimal values of $u_i^*$ in advance, and so they are defined adaptively and sequentially. First, a moderate number $n$ of Monte Carlo samples distributed according to $f(s)$ is generated. Based on these samples, $u_1^*$ is defined such that $P(F_1) = p_0$, where $p_0 \in (0,1)$ is some fixed value (typically, $p_0 = 0.1$). Next, MCMC samples distributed according to the conditional PDF $f(s|F_1)$ are generated using MMA. Based on these samples, $u_2^*$ is defined such that $P(F_2|F_1) = p_0$. The algorithm proceeds in this way until the target failure domain $F$ has been sampled sufficiently. The Subset Simulation estimate of the network failure probability $p_F$ is then given by

$$\hat{p}_F^{SS} = p_0^{L-1} \frac{n_F}{n}, \qquad (6)$$

where $L$ is number of levels, i.e. the number of intermediate failure domains in (3), $n$ is the number of generated samples at each level, and $n_F$ is the number of samples in $F$ at the last level.

The efficiency of Subset Simulation for network reliability estimation is demonstrated in (Zuev et al. 2015), where two small-world network models are compared in terms of the maximum-flow reliability of the networks they generate. For more details on and general introduction to Subset Simulation and its applications to rare event estimation, the reader is referred to the original paper, where the method was developed (Au and Beck 2001), a detailed exposition at an introductory level with implementation in MATLAB (Zuev 2015), and a fundamental monograph (Au and Wang 2014).

## 4. CHALLENGES

Estimating reliability of critical infrastructure networks is one of the central problems of modern civil engineering, operations research and applied statistics. Yet we are at the very beginning of developing efficient solutions, since there exist several fundamental challenges that make this problem both conceptually and computationally very difficult. In this section, we highlight some of these challenges and propose ways to overcome them.

In the IEEE report "Reliability analysis of complex network systems: research and practice in need" (Zio 2007), Prof. Zio observed that "…the classical methods of reliability and risk analysis fail to provide the proper instruments of analysis." A moment of thought shows that this observation is directly applicable to Subset Simulation, which is a very efficient classical reliability method, originally developed for structures (Au and Beck 2001), not networks. As explained in Section 3, to extend Subset Simulation to networks, one needs to assume that components $s_1, \ldots, s_m$ of the network state $s$ are independent, which is, of course, an oversimplification, since in real infrastructure networks $s_1, \ldots, s_m$ are correlated: a link failure often increases the probability of failure of neighboring links. Moreover, many infrastructures are prone to cascading failures, where a local failure propagates through the network and leads to a global failure (Dueñas-Osorio and Vemuru 2009; Hines et al. 2009). This shows that Subset Simulation solves the network reliability problem only approximately, and to make it useful for estimating the reliability of real infrastructure networks, realistic models of link correlations and cascading failures must be incorporated into the framework.

Over the last two decades, many different models of cascading failures in complex networks as well as methods for assessing the criticality of network components with respect to cascading failures have been proposed by researchers from different research communities, e.g. sociology (Watts 2002), physics (Crucitti et al. 2004), applied mathematics (Swift 2008), computer science (Iyear et al. 2009), and engineering (Zio and Sansavini 2011; Hernandez-Fajardo and Dueñas-Osorio 2013; Song et al. 2016), to mention but a few. A qualitatively realistic modeling of the effects of cascading failures on an infrastructure network is challenging because of 1) the complexity and variety of the physical and engineering mechanisms involved and 2) the interactions between physics of cascading failures and the decisions made by humans who operate the network. Therefore, a thorough analysis and domain-specific synthesis of developed models of cascading failures as well as modeling of "soft" factors (such as human errors, impact of online social networks, etc.) is required for accurate estimation of critical infrastructure reliability.

Another challenge in assessing reliability and resilience of critical infrastructure networks comes from the fact that they are strongly interconnected, mutually dependent, and subject to complex feedback loops. In Section 1, we discussed that the network of natural gas pipelines and the electric power grid are coupled (see Fig. 2). This cross-sector interdependence is not an exception, it is rather a rule: essentially all critical infrastructures are interconnected (Rinaldi 2001). Examples of infrastructure interdependencies are demonstrated in Figure 3. As a result, failures or poor performance of one infrastructure network can have a negative effect on other infrastructures, and this may result into a big shock to the national or even global economy. Moreover, these multi-sector interdependencies make infrastructure networks more fragile and vulnerable to cascading failures triggered by both random failures and intentional attacks, than the single networks taken in isolation (Buldyrev, 2010). These mechanisms of cascading failures are critical for a realistic risk assessment. Consequences of initial failure scenarios evolve through cascades between networks finally resulting in an economic loss or loss of human life. Since these cascades cannot be captured properly with the current network analysis approaches risk assessment is generally lacking a substantiated quantification of failure consequences.

Taking cross-sector interdependencies into account is thus absolutely necessary for both estimation of the reliability and enhancement of the resilience of critical infrastructures as well as for optimal design of interdependent networks (González et al. 2016). A survey of different approaches to modeling critical infrastructure interdependency is provided in (Pederson et al. 2006). More recently, a new methodology, based on the so-called multi-layer networks (Boccaletti et al. 2014; Kivelä et al. 2014), has been developed for modeling dependence and interactions between different complex networks and various processes on the corresponding "networks of networks" (D'Agostino and Scala 2014). Multi-layer networks have been successfully used for analyzing complex networks and extracting useful insights about the underlying complex systems in different fields of sciences and engineering.

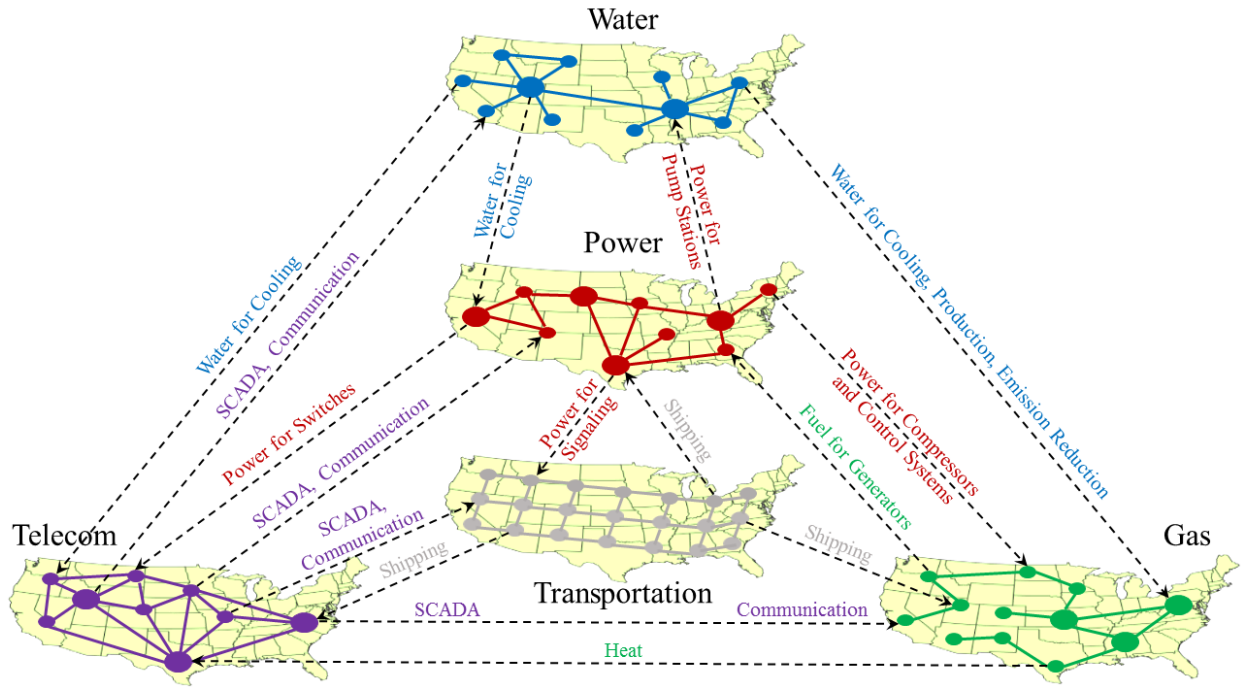

**Figure 3. Multi-sector interdependencies of critical infrastructure networks.**

This brings us to the last point of this paper: interdisciplinary collaboration is the key for understanding how to assess the reliability of infrastructure and how to make it more resilient. Complex networks are intrinsically multidisciplinary objects and they are studied by very diverse and heterogeneous research communities. For example, networks are used to analyze the spread of epidemics in human networks (Newman 2002), for predicting a financial crisis (Elliott et al. 2014), and for developing a theory of quantum gravity (Boguñá et al. 2014). Also networks may help us to understand how a brain works (Krioukov 2014) and how to treat cancer (Barabási et al. 2011). Moreover, insights obtained in network studies are often truly interdisciplinary and results obtained, for instance, for biological networks, could be extended and applied to infrastructure networks, which could be very important for making critical infrastructures more reliable and resilient. Despite the fact that the value of interdisciplinary research has been long recognized and promoted (NSF 2016), interdisciplinary collaboration in the area of critical infrastructure networks is still in its infancy. This is mainly because the area itself is relatively new: for example, prior to the 1990s, little attention was given to infrastructure interdependencies (Visarraga 2011). So it is important to organize more interdisciplinary workshops and conferences, where researchers from different fields can discuss the progress they have made, exchange ideas, describe the open problems they are facing, and, most importantly, form interdisciplinary groups for attacking these problems. Another issue concerns the conservatism of our publication media, which are typically focused on single-discipline developments. Cross-disciplinary developments are often considered as not fitting into the scope of a journal because of elements from a second discipline. We need to break off this silo structure in order to facilitate interdisciplinary discussions on this new generation of developments to make substantial progress. Similarly, research foundations and stakeholders need to translate their initial opening for interdisciplinary research into a comprehensive implementation of a interdisciplinary assessment and funding structure with interdisciplinary units, panels, etc.

It is important to realize however that this interdisciplinary "dialog" will not be smooth and easy due to substantial differences in educational backgrounds and "vocabularies" used by researcher from different fields. As illustration, Adam at al. (2015) developed a theoretical approach for studying cascading effects, which can be potentially very useful for understanding cascading failures in infrastructure networks. Understanding this paper, however, requires familiarity with notions of topological closer, isomorphism, with elements of universal algebra, and other notions which are far beyond the standard mathematical education of civil engineers. To use the results of this paper, engineers have to either educate themselves, or convince the authors to explain their results in the language engineers can understand. Neither of these two tasks is easy, but we believe both are necessary.

## 5. CONCLUSION

This paper adopts a network perspective to highlight some of the main challenges researchers face when estimating reliability of critical infrastructures. In particular, we discuss the extension of Subset Simulation, an efficient classical reliability method for estimating small failure probabilities of structures, to networks. We show that for accurate network reliability estimation, several important challenges must be overcome, including realistic domain-specific models for link correlations and cascading failures, large-scale modeling of cross-sector interdependencies and feedback loops, modeling of "soft" factors, such as human errors and interaction of infrastructure with online social networks. Domain knowledge and expertise from multiple branches of engineering, system analysis, applied mathematics, statistical physics, computer science, and computational statistics are required to successfully address the outlined challenges. To succeed in these tasks, interdisciplinary collaboration is thus a must.

## ACKNOWLEDGMENTS

The first author would like to thank the organizers of the ASCE Workshop "Resiliency of Urban Tunnels and Pipelines" for inviting him to give a keynote talk. Both authors thank the participants of the workshop for fruitful discussions and valuable suggestions.